\documentclass[pdflatex,sn-chicago]{sn-jnl}

\usepackage[utf8]{inputenc}
\usepackage{geometry}
\usepackage{graphicx}
\usepackage{booktabs}
\usepackage{tabularx}
\usepackage{multirow}
\usepackage{array}
\usepackage{amsmath,amssymb,amsfonts}
\usepackage{amsthm}
\usepackage{bm}
\usepackage{xcolor}
\usepackage{placeins}

\hypersetup{
	colorlinks=true,
	linkcolor=blue,
	citecolor=blue,
	urlcolor=blue
}
\newtheorem{theorem}{Theorem}[section]

\numberwithin{equation}{section}
\graphicspath{{./}}

\begin{document}
	
	\title[The UGARCH Model]
	{ Analysis of Uncertain GARCH Model with the application of crude oil future prices}
	
	\author{\fnm{Zhihong} \sur{Yin}}
	
	\author*{\fnm{Zhiming} \sur{Li}\textsuperscript{*}}
	\email{zmli@xju.edu.cn}
	
	\affil*{\orgdiv{College of Mathematics and System Science},
		\orgname{Xinjiang University},
		\orgaddress{\city{Urumqi}, \country{China}}}

	\abstract {Time series data often exhibit variation, clustering, and persistence in uncertain environments. Based on an uncertain measure, existing studies have primarily focused on conditional mean structures while ignoring variance and persistent dynamics. To address this gap, this paper develops an uncertain generalized autoregressive conditional heteroskedasticity (UGARCH) model that incorporates lagged conditional variance terms. We first prove sufficient conditions to ensure mean stationarity of the variance process. Then, we propose recursive constrained least squares to estimate the model parameters and use cross-validation to determine the orders of the mean and variance equations. In addition, we also employ uncertain hypothesis tests to determine whether a time series exhibits a UGARCH effect and to assess the goodness of fit. Point and interval forecasts are subsequently constructed from the recursively obtained next-period conditional variance and the expected value of the mean equation. Numerical simulations demonstrate the parameter estimation performance of the proposed method and show that UGARCH achieves smaller conditional-variance fitting errors than UARCH under the considered settings. In the empirical analysis, we apply the proposed method to West Texas Intermediate crude oil futures closing prices, which exhibit time-varying volatility and support short-term point and interval forecasts.}
	
	\keywords{Uncertain measure, Uncertain time-series analysis, UGARCH model,   hypothesis test}
	
	\maketitle

	\section{Introduction}\label{sec:introduction}	 
	
	Time-series analysis is an important tool for describing the dynamic
	dependence of chronologically ordered observations and forecasting
	their future values. Early studies mainly focused on conditional mean
	dynamics. \citet{yule1927} introduced the autoregressive (AR) model,
	\citet{walker1931} studied the moving average (MA) model, and
	\citet{boxJenkins1970} established a systematic framework for
	autoregressive moving average (ARMA) analysis. However, financial
	time series often exhibit volatility clustering and persistence,
	which cannot be adequately characterized by conditional mean models
	alone. To address time-varying volatility, \citet{engle1982}
	proposed the autoregressive conditional heteroskedasticity (ARCH)
	model, in which conditional variance depends on past squared
	disturbances. \citet{bollerslev1986} further introduced lagged
	conditional variances and developed the generalized autoregressive
	conditional heteroskedasticity (GARCH) model, providing a relatively
	parsimonious framework for capturing volatility persistence.
	Subsequently, \citet{nelson1991} proposed the exponential GARCH
	(EGARCH) model, which models the logarithm of conditional variance
	and allows asymmetric responses to positive and negative shocks.
	\citet{glosten1993} further investigated asymmetric volatility
	effects by allowing shocks of different signs to have different
	impacts on conditional variance. \citet{ding1993} examined the
	long-memory properties of financial volatility and developed a
	power-based conditional heteroskedasticity model, while
	\citet{engleLee1999} distinguished between long-run and short-run
	components of volatility. More recently, \citet{conradKleen2020} investigated multiplicative component GARCH-MIDAS models for volatility forecasting. \citet{verma2021} combined GARCH-type models
	with recurrent neural networks to forecast crude oil
	futures volatility. \citet{salisuGuptaDemirer2022}
	employed the GARCH-MIDAS framework to examine the
	predictive role of the global financial cycle in oil
	market volatility. \citet{wangMaLiuLang2023} extended
	the GARCH-MIDAS framework by incorporating extreme shocks
	to improve oil volatility forecasting. \citet{huangLuo2024}
	studied hybrid GARCH-type models incorporating long memory,
	regime switching, leverage effects, and heavy tails. \citet{yanKellardLambercy2025}
	developed multivariate range-based EGARCH models for covariance
	forecasting. \citet{nikolakopoulos2025} proposed a Bayesian
	semiparametric multivariate realized GARCH framework for modeling
	returns and realized covariance. 
	 \citet{archakovHansenLunde2026} developed a multivariate
	realized GARCH model incorporating realized measures of volatility
	and correlations.  \citet{fiszederMalecka2026}
	proposed robust multivariate range-based GARCH models and examined
	their forecasting performance across oil, gold, equity, and Bitcoin
	markets.
	
	The above models are primarily based on probability theory. However, when observations are limited or imprecise, sufficient frequency information may not be available
	to determine an appropriate probability distribution. To describe
	uncertain phenomena under such circumstances, \citet{liu2007}
	introduced uncertainty theory,  further developed by
	\citet{liu2009} and \citet{liu2010}. Based on this theoretical
	framework, \citet{yangLiu2019} established the uncertain
	autoregressive (UAR) model and investigated parameter estimation,
	residual analysis, and forecasting methods. \citet{zhao2020} derived an analytic least-squares solution
	for the UAR model.
	\citet{yangParkHu2020} proposed a least absolute deviations
	estimation method for UAR parameters.
	\citet{chenYang2021} investigated maximum likelihood estimation
	for the UAR model.
	\citet{liuYang2022} developed a cross-validation approach
	for selecting the order of the UAR model.To characterize
	the effects of historical uncertain disturbances,
	\citet{yangNi2021} proposed the uncertain moving average (UMA)
	model, which was further extended to higher-order cases by
	\citet{xinYangGao2021}. \citet{xin2023} subsequently combined
	autoregressive and moving average terms to establish the uncertain
	autoregressive moving average (UARMA) model, while
	\citet{liuLi2024} investigated its least absolute deviations
	estimation method. These studies have progressively enriched
	the modeling and estimation methods for uncertain time series,
	but have mainly focused on the dynamics of the observed series
	rather than explicitly modeling time-varying disturbance variances.
	
	Recently, volatility modeling under uncertainty theory has
	attracted increasing attention. \citet{shiWangLiang2026}
	developed uncertain HAR-RV models for forecasting crude oil
	futures volatility. \citet{gaoChen2026} employed uncertain
	vector autoregression to jointly forecast oil and equity
	market volatility. \citet{wangShiLiang2026} proposed uncertain
	threshold HAR models for WTI crude oil futures.
	\citet{han2026} introduced the uncertain autoregressive
	conditional heteroskedasticity (UARCH) model based on lagged
	squared centered residuals. However, the UARCH model does not explicitly incorporate lagged conditional variances and may require higher lag orders to capture strong volatility persistence. { To address this limitation, the main contributions are summarized as follows:
\begin{itemize}
\item[(i)] Propose an uncertain generalized autoregressive conditional heteroskedasticity (UGARCH) model by introducing lagged conditional variances into the UARCH model. The proposed model captures both past volatility shocks and variance persistence, and reduces to the centered UARCH model when the lagged conditional variance coefficients are zero.
\item[(ii)] Prove a sufficient condition for the asymptotic mean stability of the conditional variance process and develop a recursive constrained least-squares estimation method. Cross-validation and uncertain hypothesis tests are employed for order selection and model assessment, respectively. 
\item[(iii)] Conduct numerical simulations to evaluate the estimation and fitting performance of the proposed model, and apply to WTI crude oil futures closing prices for time-varying volatility analysis and short-term point and interval forecasting.
\end{itemize}	
}

The remainder of this paper is organized as follows.
	Section~\ref{sec2} reviews the fundamental concepts of
	uncertainty theory and introduces the UAR and UGARCH models.
	Section~\ref{sec3} presents the stability analysis, parameter
	estimation, model order selection, uncertain hypothesis testing,
	and forecasting procedure.
	Section~\ref{sec4} presents numerical simulations to evaluate
	the proposed model.
	Section~\ref{sec5} provides an empirical analysis using WTI
	crude oil futures prices, and Section~\ref{sec6} concludes
	the paper.

\section{Preliminaries and the UGARCH model}\label{sec2}  

In this section, we first review the fundamental concepts of uncertainty theory  
following the work of \citet{liu2026}, including uncertain measure, uncertain variable,  
uncertainty distribution and its inverse, expected value, and variance, and then  
establish the UGARCH model for time series.

Let $\Gamma$ be a nonempty set and $\mathcal L$ be a $\sigma$-algebra over $\Gamma$. A set function $\mathcal M:\mathcal L\rightarrow[0,1]$ is called an \emph{uncertain measure} if it satisfies the following three axioms: (i) normality: $\mathcal M\{\Gamma\}=1$, (ii) duality: $\mathcal M\{\Lambda\}+\mathcal M\{\Lambda^{\mathrm c}\}=1$ for every event $\Lambda\in\mathcal L$, and (iii) subadditivity: for every countable sequence of events $\Lambda_1,\Lambda_2,\ldots$, $\mathcal M \{\bigcup_{i=1}^{\infty}\Lambda_i\}			\leq\sum_{i=1}^{\infty}\mathcal M\{\Lambda_i\}.$ The triplet $(\Gamma,\mathcal L,\mathcal M)$ is called an \emph{uncertainty space}.  Under the uncertainty space $(\Gamma,\mathcal L,\mathcal M)$, a function $\xi:\Gamma\rightarrow\mathbb R$ is called an \emph{uncertain variable} if $\{\xi\in B\}=\{\gamma\in\Gamma:\xi(\gamma)\in B\}	$ is an event for every Borel set $B$ of real numbers.  
The function $\Phi(x)=\mathcal M\{\xi\leq x\}( x\in\mathbb R)$ is called the \emph{uncertainty distribution} of the uncertain variable $\xi$.  If $\Phi$ is continuous, strictly increasing at every $x$ for which $0<\Phi(x)<1$, and satisfies $\lim_{x\rightarrow-\infty}\Phi(x)=0,$ and $\lim_{x\rightarrow+\infty}\Phi(x)=1,$  
then $\Phi$ is called a \emph{regular uncertainty distribution}. For $0<\alpha<1$,  the inverse function $\Phi^{-1}(\alpha)$ is called the \emph{inverse uncertainty distribution} of $\xi$.  The \emph{expected value} of $\xi$ is defined as  
\[  
E[\xi]=\int_{0}^{+\infty}\mathcal M\{\xi\geq x\}\,\mathrm dx  
-\int_{-\infty}^{0}\mathcal M\{\xi\leq x\}\,\mathrm dx  
\]  
if at least one of the two integrals is finite. If $\xi$ has a finite expected value $e$, then  
$  
V[\xi]=E\left[(\xi-e)^2\right]  
$  
is called its \emph{variance}. If $\xi$ has a regular uncertainty distribution $\Phi$, then  
\[  
E[\xi]=\int_{0}^{1}\Phi^{-1}(\alpha)\,\mathrm d\alpha, \quad V[\xi]=\int_{0}^{1}\left[\Phi^{-1}(\alpha)-e\right]^2\,\mathrm d\alpha.  
\]  

An uncertain variable $\xi$ is normal if it has the uncertainty distribution  
\[  
\Phi(x)=\left[1+\exp\left(\frac{\pi(e-x)}{\sqrt{3}\sigma}\right)\right]^{-1},  
\qquad x\in\mathbb R,  
\]  
where $e$ and $\sigma$ are real numbers and $\sigma>0$, denoted by $\xi\sim N(e,\sigma)$. 	Thus, $E[\xi]=e$ and $V[\xi]=\sigma^2.$	For $0<\alpha<1$,  its inverse uncertainty distribution is  
$\Phi^{-1}(\alpha)=e+\frac{\sqrt{3}\sigma}{\pi}\ln\frac{\alpha}{1-\alpha}.	$  

Based on the above definitions, we next describe the modeling process of the UGARCH model via an uncertain autoregressive (UAR) model.  Let $X_1, X_2,\ldots, X_m$ be the observed values of a time series at times $1,2,\ldots,m$, respectively. \citet{yangLiu2019} proposed a UAR model of order $p$, formed by  
\begin{align}\label{UAR}  
	X_t=b_0+\sum_{j=1}^{p}b_jX_{t-j}+\epsilon_t,  
	\qquad t=p+1,p+2,\ldots,m,  
\end{align}  
where $b_0,b_1,\ldots,b_p$ are unknown parameters, and $\epsilon_t$ is the uncertain disturbance at time $t$ with $E[\epsilon_t]=e$ and $V[\epsilon_t]=\sigma^2$.  Let $\hat b_0, \hat b_1, \ldots, \hat b_p$ be the estimators of $b_0,b_1,\ldots,b_p$. By the least-squares method, we have   
\begin{equation*}  
	(\hat b_0, \hat b_1, \ldots, \hat b_p)  
	=  
	\underset{b_0,b_1,\ldots,b_p}{\arg\min}  
	\sum_{t=p+1}^{m}  
	\left(  
	X_t  
	-b_0  
	-\sum_{j=1}^{p} b_j X_{t-j}  
	\right)^2.  
\end{equation*}  

The fitted value of $X_t$ obtained from the UAR (\ref{UAR}) is  
$\hat b_0+\sum_{j=1}^{p}\hat b_jX_{t-j}.$  
The model residual at time $t$ is therefore defined as  
$\hat{\epsilon}_t=X_t-\hat b_0-\sum_{j=1}^{p}\hat b_jX_{t-j}$ for $t=p+1,p+2,\ldots,m.$  
These residuals represent the variation that remains unexplained after removing the autoregressive mean structure and form the basis of the subsequent volatility series. From the residual sample, the expected value $e$ is estimated by $\hat e=\frac{1}{m-p}\sum_{t=p+1}^{m}\hat\epsilon_t,$ and the standard deviation $\sigma$ is estimated by $\hat\sigma=\sqrt{\frac{1}{m-p}\sum_{t=p+1}^{m}(\hat\epsilon_t-\hat e)^2}.$  

We assume that $\epsilon_{m+1}\sim N(\hat e,\hat\sigma)$. Then, conditional on the historical observations, the one-step-ahead forecast uncertain variable $X_{m+1}$ is represented by 
\begin{equation*}   
	X_{m+1}=\hat b_0+\sum_{j=1}^{p}\hat b_jX_{m+1-j}   
	+N(\hat e,\hat\sigma).   
\end{equation*}
Its expected value provides a point forecast:   
$\hat\mu_{m+1}=\hat b_0+\sum_{j=1}^{p}\hat b_jX_{m+1-j}+\hat e.$  
Conditional on the historical observations and treating the estimated model parameters as fixed, we have  
\begin{equation*}
	V[X_{m+1}]
	= E[(X_{m+1}-\hat\mu_{m+1})^2]
	= E[(\epsilon_{m+1}-\hat e)^2]
	= V[\epsilon_{m+1}].
\end{equation*}

These results show that the UAR model (\ref{UAR}) characterizes the conditional mean of a time series, whereas the variance of the forecast uncertain variable is determined by the disturbance variance. Generally, existing UAR analyses assume a constant disturbance variance and therefore cannot describe time-varying volatility and volatility clustering in financial time series.   

To capture time variation in the disturbance variance, the UARCH model, proposed by \citet{han2026}, describes time-varying volatility using lagged squared centered residuals. Based on the residuals obtained from the fitted UAR model, they defined a squared-centered residual series $h_t=(\hat{\epsilon}_t-\hat e)^2$, $t=p+1,p+2,\ldots,m$, and modeled its dynamics with an autoregressive structure. Thus, the UARCH($q$) model is specified as   
\begin{align}\label{uarch}h_t=\alpha_0+\sum_{i=1}^{q}\alpha_ih_{t-i}+\eta_t,\quad t=p+q+1,p+q+2,\ldots,m,  
\end{align}  
where $\alpha_0>0$, $\alpha_i\geq 0$ $(i=1,\ldots,q)$, and $\eta_t$ is an uncertain disturbance term with $E[\eta_t]=\theta$ and $V[\eta_t]=\omega^2$. Thus, the UARCH model allows the disturbance variance to evolve over time based on past squared centered residuals, rather than assuming a constant disturbance variance.  

However, its variance equation contains only past squared centered residuals; consequently, representing volatility persistence generally requires increasing the model order, which complicates parameter estimation. To address this limitation, we introduce lagged conditional variances into the UARCH variance equation to capture the persistent effect of past volatility shocks. In the proposed UGARCH model, we use a centered uncertain disturbance term. If $E[\eta_t]=\theta$, then
$\alpha_0+\eta_t=(\alpha_0+\theta)+(\eta_t-\theta)$,
so the disturbance mean can be absorbed into the intercept without changing its variance. After this reparameterization, we continue to denote the intercept by $\alpha_0$ and the centered disturbance by $\eta_t$ for simplicity. Hence, $E[\eta_t]=0$ and $V[\eta_t]=\omega^2$ in the UGARCH model.

This yields the UGARCH$(q,s)$ model. Let \(\sigma_t^2\) be the conditional variance at time \(t\), in the sense that it represents the time-varying variance level determined from the historical information. Denote \(r=\max(q,s)\), which is the largest lag order appearing in the variance equation. For $t=p+r+1,\ldots,m$, the UGARCH$(q,s)$ model is established as follows
\begin{equation}\label{eq:ugarch}  
	h_t=\sigma_t^2+\eta_t, \quad  
	\sigma_t^2=\alpha_0+\sum_{i=1}^{q}\alpha_i h_{t-i}  
	+\sum_{j=1}^{s}\beta_j\sigma_{t-j}^2,  
\end{equation}  
where \(q\) and \(s\) are the lag orders of the squared centered residuals  
and conditional variances, respectively, and \(\eta_t\) is a centered uncertain  
disturbance satisfying $E[\eta_t]=0,$ $V[\eta_t]=\omega^2.$   To ensure that the recursively computed conditional variances are  
nonnegative, we impose  
$\alpha_0>0, \alpha_i\geq0( i=1,\ldots,q), \beta_j\geq0( j=1,\ldots,s).$  
When all \(\beta_j=0\), the UGARCH$(q,s)$ model  
reduces to a centered form of the UARCH$(q)$ model.

\section{Uncertain statistical analysis}\label{sec3}
This section focuses on uncertain statistical analysis of the UGARCH model (\ref{eq:ugarch}), including mean stability, parameter estimation, order selection, hypothesis testing, and forecasting.
\subsection{Mean stability}	 

The stability of the conditional variance process matters for volatility modeling because it ensures that past volatility shocks do not accumulate indefinitely and that variance remains finite in the long run. In classical GARCH models, stationarity guarantees a finite long-run variance and implies mean-reverting volatility behavior (Bollerslev (1986)). Under an uncertain measure, Han et al. (2026) studied the stationarity of the UARCH model to ensure stable long-run behavior of the squared centered residual process. In this subsection, we study the mean stability of the conditional variance process in the UGARCH model (\ref{eq:ugarch}).

\begin{theorem}
	In the UGARCH$(q,s)$ model (\ref{eq:ugarch}),
	denote $r=\max(q,s)$ and $c_l=\alpha_l+\beta_l$
	for $l=1,\ldots,r$, where $\alpha_l=0$ for
	$l>q$ and $\beta_l=0$ for $l>s$.
	If all roots $z$ of the characteristic equation
	$1-c_1z-c_2z^2-\cdots-c_rz^r=0$
	have absolute values greater than $1$,
	or equivalently,
	$\sum_{i=1}^{q}\alpha_i+\sum_{j=1}^{s}\beta_j<1$,
	then the conditional variance process is mean-stationary.
\end{theorem}

\begin{proof}
	For the UGARCH$(q,s)$ model (\ref{eq:ugarch}), taking expectations on both sides of the second equation gives
	\[
	\operatorname E[\sigma_t^2]
	=\alpha_0+\sum_{i=1}^{q}\alpha_i\operatorname E[h_{t-i}]
	+\sum_{j=1}^{s}\beta_j\operatorname E[\sigma_{t-j}^2].
	\]
	Since $h_{t-i}=\sigma_{t-i}^2+\eta_{t-i}$ and $\operatorname E[\eta_{t-i}]=0$, we have
	$\operatorname E[h_{t-i}]=\operatorname E[\sigma_{t-i}^2]$ $(i=1,2,\ldots,q)$. Thus, it follows that
	\[
	\operatorname E[\sigma_t^2]
	=\alpha_0+\sum_{i=1}^{q}\alpha_i\operatorname E[\sigma_{t-i}^2]
	+\sum_{j=1}^{s}\beta_j\operatorname E[\sigma_{t-j}^2]
	=\alpha_0+\sum_{l=1}^{r}c_l\operatorname E[\sigma_{t-l}^2].
	\]
	Denote $\boldsymbol\Sigma_t=(\sigma_t^2,\sigma_{t-1}^2,\ldots,\sigma_{t-r+1}^2)^T$, $\boldsymbol a=(\alpha_0,0,\ldots,0)^T$, and
	\[
	\boldsymbol C=
	\begin{pmatrix}
		c_1&c_2&\cdots&c_{r-1}&c_r\\
		1&0&\cdots&0&0\\
		0&1&\cdots&0&0\\
		\vdots&\vdots&\ddots&\vdots&\vdots\\
		0&0&\cdots&1&0
	\end{pmatrix}.
	\]
	The preceding recursion can then be written as
	$\operatorname E[\boldsymbol\Sigma_t]
	=\boldsymbol a+\boldsymbol C\operatorname E[\boldsymbol\Sigma_{t-1}]$.
	Repeated iteration yields
	\[
	\operatorname E[\boldsymbol\Sigma_t]
	=\left(I+\boldsymbol C+\boldsymbol C^2+\cdots+\boldsymbol C^{n-1}\right)\boldsymbol a
	+\boldsymbol C^n\operatorname E[\boldsymbol\Sigma_{t-n}].
	\]
	By construction, the characteristic polynomial of $\boldsymbol C$ is
	$|\lambda I-\boldsymbol C|
	=\lambda^r-c_1\lambda^{r-1}-c_2\lambda^{r-2}-\cdots-c_r$.
	Setting $\lambda=1/z$ and multiplying both sides by $z^r$ shows that $\lambda$ is an eigenvalue of $\boldsymbol C$ if and only if $z=1/\lambda$ is a root of the stated characteristic equation. By assumption, every root satisfies $|z|>1$, so every eigenvalue of $\boldsymbol C$ satisfies $|\lambda|=1/|z|<1$. Hence,
	$\lim_{n\to\infty}\boldsymbol C^n=\boldsymbol0$, and
	$\lim_{n\to\infty}(I+\boldsymbol C+\cdots+\boldsymbol C^{n-1})=(I-\boldsymbol C)^{-1}$.
	Letting $n\to\infty$ gives
	$\operatorname E[\boldsymbol\Sigma_t]\longrightarrow(I-\boldsymbol C)^{-1}\boldsymbol a$.
	Thus, $\operatorname E[\boldsymbol\Sigma_t]$ converges to a finite vector, whose first component is the expected conditional variance $\operatorname E[\sigma_t^2]$. In particular, the long-run expected conditional variance is
	\[
	\lim_{t\to\infty}\operatorname E[\sigma_t^2]
	=\frac{\alpha_0}
	{1-\sum_{i=1}^{q}\alpha_i-\sum_{j=1}^{s}\beta_j}.
	\]
	Therefore, the conditional variance process of the UGARCH$(q,s)$ model (\ref{eq:ugarch}) is mean stationary.
\end{proof}

We next show that the root condition in Theorem 3.1 is equivalent to $\sum_{i=1}^{q}\alpha_i+\sum_{j=1}^{s}\beta_j<1$.
Let $P(z)=1-c_1z-c_2z^2-\cdots-c_rz^r$ and $S=\sum_{l=1}^{r}c_l$.Suppose that all roots $z$ of the characteristic equation satisfy $|z|>1$. Then, $P(x)$ has no zeros on the real interval $[0,1]$. Since $P(x)$ is continuous and $P(0)=1>0$, it follows that $P(x)>0$ in $[0,1]$. In particular,
$P(1)=1-\sum_{l=1}^{r}c_l=1-S>0$, and therefore $S<1$.
Conversely, suppose that $S<1$. If $|z|\leq1$, then
\[
\left|\sum_{l=1}^{r}c_lz^l\right|
\leq\sum_{l=1}^{r}c_l|z|^l
\leq\sum_{l=1}^{r}c_l=S<1.
\]
Thus, $\sum_{l=1}^{r}c_lz^l\neq1$, or equivalently $P(z)\neq0$. Then, all roots of $P(z)=0$ satisfy $|z|>1$.
	
	\subsection{Parameter estimation} 
	
	Unlike the UARCH model (\ref{uarch}), the UGARCH model (\ref{eq:ugarch}) contains lagged 
	conditional variances \(\sigma_{t-j}^2\), which are not directly observable 
	and must be recursively computed from the model parameters. Consequently, 
	UGARCH parameter estimation cannot be reduced to a single ordinary linear 
	least-squares problem. We will use recursive constrained least squares 
	to estimate the model parameters through the following steps. 
	
	Let 	$\boldsymbol\vartheta 
	= 
	(\alpha_0,\alpha_1,\ldots,\alpha_q, 
	\beta_1,\ldots,\beta_s)^\top$ and  
	$n=m-p-r.$ 
	To ensure nonnegative recursively computed conditional variances and model 
	stability, let \(\varepsilon(>0)\) be a sufficiently small constant and define 
	the feasible parameter space 
	\[ 
	\Theta_\varepsilon 
	= 
	\left \{ 
	\boldsymbol\vartheta: 
	\alpha_0\geq\varepsilon,\  
	\alpha_i\geq0,\  
	\beta_j\geq0,\  
	\sum_{i=1}^{q}\alpha_i+\sum_{j=1}^{s}\beta_j 
	\leq1-\varepsilon 
	\right\}. 
	\] 
	
	Using the squared centered residuals $h_t$ defined in
	Section 2, we estimate the UGARCH parameters by the
	following four steps: 
	
	\textit{Step i. Parameter initialization.}  	{ For \(s\geq1\), let 
		$\beta_1^{(0)}=\cdots=\beta_s^{(0)}=0$, so that the lagged conditional variance terms are excluded from the initial variance equation. When \(s=0\), no \(\beta_j\) parameters are present. In both cases,} 
	the initial variance equation reduces to a UARCH$(q)$ model (\ref{uarch}). 
	We solve 
	\[ 
	(\alpha_0^{(0)},\ldots,\alpha_q^{(0)}) 
	= 
	\underset{\alpha_0,\ldots,\alpha_q}{\arg\min} 
	\sum_{t=p+r+1}^{m} 
	(h_t-\alpha_0 
	-\sum_{i=1}^{q}\alpha_i h_{t-i})^2, 
	\] 
	subject to 	$\alpha_0\geq\varepsilon, \alpha_i\geq0, 	\sum_{i=1}^{q}\alpha_i\leq1-\varepsilon.$  
	This gives the initial parameter vector 
	$\boldsymbol\vartheta^{(0)}=(\alpha_0^{(0)},\alpha_1^{(0)},\ldots,	\alpha_q^{(0)},0,\ldots,0)^\top$. 
	
	\textit{Step ii. Recursive computation of conditional variances.} 	Given the parameter vector at iteration \(k\), 
	$\boldsymbol\vartheta^{(k)}=(\alpha_0^{(k)},\alpha_1^{(k)},\ldots, 	\alpha_q^{(k)},\beta_1^{(k)},\ldots,\beta_s^{(k)})^\top,$ 	initialize the conditional variances by 
	$ 
	(\hat\sigma_t^2)^{(k)}= h_t$  for 
	$t=p+1,\ldots,p+r. 
	$ 
	Then, for \(t=p+r+1,\ldots,m\), we recursively compute 
	\[ 
	(\hat\sigma_t^2)^{(k)}
	= 
	\alpha_0^{(k)} 
	+\sum_{i=1}^{q}\alpha_i^{(k)} h_{t-i} 
	+\sum_{j=1}^{s}\beta_j^{(k)}(\hat\sigma_{t-j}^2)^{(k)}. 
	\] 
	
	\textit{Step iii. Parameter update.} Holding the conditional variance path obtained at iteration \(k\) fixed, 
	define 
	$ 
	\boldsymbol Y 
	= 
	( h_{p+r+1}, h_{p+r+2},\ldots, h_m)^\top. 
	$ 
	The corresponding design matrix is 
	\[ 
	\boldsymbol Z^{(k)} 
	= 
	\begin{pmatrix} 
		1& h_{p+r}&\cdots& h_{p+r+1-q}& 
		(\hat\sigma_{p+r}^2)^{(k)}&\cdots& 
		(\hat\sigma_{p+r+1-s}^2)^{(k)}\\ 
		1& h_{p+r+1}&\cdots& h_{p+r+2-q}& 
		(\hat\sigma_{p+r+1}^2)^{(k)}&\cdots& 
		(\hat\sigma_{p+r+2-s}^2)^{(k)}\\ 
		\vdots&\vdots&&\vdots&\vdots&&\vdots\\ 
		1& h_{m-1}&\cdots& h_{m-q}& 
		(\hat\sigma_{m-1}^2)^{(k)}&\cdots& 
		(\hat\sigma_{m-s}^2)^{(k)} 
	\end{pmatrix}. 
	\] 
	When \(s=0\), the columns corresponding to lagged conditional variances are 
	omitted. The estimation equation can therefore be written as 
	$ 
	\boldsymbol Y 
	= 
	\boldsymbol Z^{(k)}\boldsymbol\vartheta+\boldsymbol\eta. 
	$ 
	The parameter estimate at iteration \(k+1\) is 
	$$ 
	\boldsymbol\vartheta^{(k+1)} 
	= 
	\underset{\boldsymbol\vartheta\in\Theta_\varepsilon}{\arg\min} 
	\|\boldsymbol Y-\boldsymbol Z^{(k)}\boldsymbol\vartheta \|_2^2, 
	$$ 
	or  equivalently, 
	\[ 
	\boldsymbol\vartheta^{(k+1)} 
	= 
	\underset{\boldsymbol\vartheta\in\Theta_\varepsilon}{\arg\min} 
	\sum_{t=p+r+1}^{m} 
	\left( 
	h_t-\alpha_0 
	-\sum_{i=1}^{q}\alpha_i h_{t-i} 
	-\sum_{j=1}^{s}\beta_j(\hat\sigma_{t-j}^2)^{(k)} 
	\right)^2. 
	\] 
	Because the conditional variance path obtained at iteration \(k\) is held 
	fixed during this parameter update, the preceding problem is a constrained 
	linear least-squares problem. 
	
	\textit{Step iv. Convergence assessment.}  
	After obtaining \(\boldsymbol\vartheta^{(k+1)}\), recompute the conditional 
	variance path as in Step 2. Given a convergence tolerance \(\delta>0\) and 
	a maximum number of iterations \(K_{\max}\), stop the iteration if 
	$ 
	\| 
	\boldsymbol\vartheta^{(k+1)} 
	-\boldsymbol\vartheta^{(k)} 
	\|_\infty 
	<\delta, 
	$ 
	and set 
	$ 
	\hat{\boldsymbol\vartheta} 
	= 
	\boldsymbol\vartheta^{(k+1)}. 
	$ 
	If the convergence criterion is not satisfied after \(K_{\max}\) 
	iterations, the algorithm is terminated, and the lack of convergence is 
	reported. 
	
	In summary, the proposed method first computes the conditional variance 
	recursively from the current parameter values, then solves a constrained 
	least-squares problem while holding the conditional variance path fixed, 
	and finally recomputes the path using the updated parameters. These steps 
	are repeated until convergence or until the maximum number of iterations 
	is reached.

	\subsection{Order selection for the UGARCH model}  
	
	Before constructing the UAR$(p)$ model (\ref{UAR}) and the UGARCH$(q,s)$ model (\ref{eq:ugarch}), the orders $p$ and $(q,s)$ must be determined. Following \citet{liuYang2022}, we use the \emph{average testing error} (ATE) to evaluate the out-of-sample predictive performance of candidate models. Specifically, model parameters are estimated from a training set, and prediction errors are then evaluated on a subsequent test set. A smaller ATE indicates better predictive performance, and the order with the smallest ATE is selected.  
	
	For the UAR$(p)$ model, we employ rolling-origin cross-validation, in which the training sample is successively expanded, and the model is re-estimated to reduce the dependence of order selection on a single forecasting origin. For the UGARCH$(q,s)$ model, since each re-estimation of the variance equation requires the complete iterative least-squares procedure, we adopt fixed-origin cross-validation to reduce the computational burden.  
	
	Let $X_1,\ldots, X_m$ be independent uncertain variables with regular uncertainty distributions $\Phi_1,\ldots,\Phi_m$. Given an initial training-sample length $T$ satisfying $p<T<m$, we estimate $\hat b_0^{(T+u)},\ldots,\hat b_p^{(T+u)}$ from the training subsample $\{X_1,\ldots,X_{T+u}\}$ at each origin $T+u$, where $u=0,1,\ldots,m-1-T$. Define the rolling-origin average testing error as  
	\begin{equation*}  
		\begin{aligned}  
			\mathrm{ATE}_T(p)  
			={}&  
			\sum_{u=0}^{m-1-T}  
			\frac{1}{m-T-u}  
			\sum_{t=T+u+1}^{m}  
			E\!\left[  
			\left(  
			X_t-\hat b_0^{(T+u)}  
			-\sum_{j=1}^{p}  
			\hat b_j^{(T+u)}X_{t-j}  
			\right)^2  
			\right].  
		\end{aligned}  
	\end{equation*}  
	By the linear operational law for uncertain variables, this quantity can be evaluated as  
	\begin{equation*}  
		\begin{aligned}  
			\mathrm{ATE}_T(p)  
			={}&  
			\sum_{u=0}^{m-1-T}  
			\frac{1}{m-T-u}  
			\sum_{t=T+u+1}^{m}  
			\int_0^1  
			\left[  
			\Phi_t^{-1}(\alpha)  
			-\hat b_0^{(T+u)}  
			-\sum_{j=1}^{p}  
			\hat b_j^{(T+u)}  
			\Upsilon_{t-j}^{-1}  
			\bigl(\alpha;\hat b_j^{(T+u)}\bigr)  
			\right]^2  
			\,\mathrm{d}\alpha,  
		\end{aligned}  
	\end{equation*}  
	where  
	\begin{equation*}  
		\Upsilon_{t-j}^{-1}(\alpha;\hat b_j^{(T+u)})=  
		\begin{cases}  
			\Phi_{t-j}^{-1}(1-\alpha),&\hat b_j^{(T+u)}\geq0,\\[4pt]  
			\Phi_{t-j}^{-1}(\alpha),&\hat b_j^{(T+u)}<0.  
		\end{cases}  
	\end{equation*}  
	The selected order is  
	$  
	\hat p=\arg\min_p\ \mathrm{ATE}_T(p).  
	$  
	
	After selecting $\hat p$, the squared centered residuals $h_{\hat p+1},\ldots,h_m$ are obtained. There are $N=m-\hat p$ observations. Letting $\tau=t-\hat p$, reindex them as $h_1,\ldots,h_N$. Given a training-sample length $T_h$ satisfying $r<T_h<N$, where $r=\max(q,s)$, consider each candidate pair $(q,s)$, with $q=1,\ldots,Q_{\max}$ and $s=0,1,\ldots,S_{\max}$. The steps of order selection are shown as follows:  
	
	\textit{Step I. Estimate parameters from the training set.} Using the training subsample $h_1,\ldots,h_{T_h}$, we apply the parameter-estimation procedure and obtain  
	$\hat\alpha_0^{(T_h)},\hat\alpha_1^{(T_h)},\ldots,\hat\alpha_q^{(T_h)}, \hat\beta_1^{(T_h)},\ldots,\hat\beta_s^{(T_h)},$ together with the fitted conditional variances $\hat\sigma_\tau^{2(T_h)}$ for $\tau=1,\ldots,T_h$.  
	
	\textit{Step II. Extend the recursion over the test period.} Holding these parameters fixed, for $\tau=T_h+1,\ldots,N$, we compute  
	\[  
	\hat\sigma_\tau^{2(T_h)}  
	=  
	\hat\alpha_0^{(T_h)}  
	+\sum_{i=1}^{q}\hat\alpha_i^{(T_h)} h_{\tau-i}  
	+\sum_{j=1}^{s}\hat\beta_j^{(T_h)}  
	\hat\sigma_{\tau-j}^{2(T_h)}.  
	\]  
	
	\textit{Step III. Compute the average testing error.}  
	Define  
	\[  
	\mathrm{ATE}_{T_h}(q,s)  
	=  
	\frac{1}{N-T_h}  
	\sum_{\tau=T_h+1}^{N}  
	(h_\tau-\hat\sigma_\tau^{2(T_h)})^2.  
	\]  
	The selected orders are  
	$  
	(\hat q,\hat s)=\arg\min_{q,s}\ \mathrm{ATE}_{T_h}(q,s).  
	$

	\subsection{Uncertain hypothesis test}
	
	The UGARCH model (\ref{eq:ugarch}) is designed to describe conditional heteroskedasticity. In applications, however, it is necessary first to determine whether the observed series exhibits conditional heteroskedasticity. After a UGARCH model (\ref{eq:ugarch}) has been fitted, a residual goodness-of-fit test is further conducted to assess the adequacy of the selected model. We first consider the hypotheses
	\[
	H_0:\alpha_1=\cdots=\alpha_q=\beta_1=\cdots=\beta_s=0
	\quad\text{versus}\quad
	H_1:\alpha_1,\ldots,\alpha_q,\beta_1,\ldots,\beta_s
	\text{ are not all zero}.
	\]
	
	Based on the fitted UAR$(p)$ model, the squared centered residuals 
	$h_t=(\hat\epsilon_t-\hat e)^2$, $t=p+1,\ldots,m$, are used to determine whether the disturbance variance is constant. Under $H_0$, the squared centered residuals are represented by the constant-variance model $h_t=\alpha_0+\eta_t$. The constant term is estimated by 
	$\hat\alpha_0=\frac{1}{m-p}\sum_{t=p+1}^{m}h_t$, and the corresponding residuals are 
	$\hat\eta_{0,t}=h_t-\hat\alpha_0$, $t=p+1,\ldots,m$. Their estimated mean and variance are
	\[
	\hat\theta_0
	=
	\frac{1}{m-p}\sum_{t=p+1}^{m}\hat\eta_{0,t},
	\qquad
	\hat\omega_0^2
	=
	\frac{1}{m-p}\sum_{t=p+1}^{m}
	(\hat\eta_{0,t}-\hat\theta_0)^2.
	\]
	We assume that the estimated disturbance term $\hat\eta_{0,t}$
	follows a normal uncertainty distribution
	$N(\hat\theta_0,\hat\omega_0)$, whose inverse uncertainty
	distribution is
	$\Phi_0^{-1}(u)=\hat\theta_0+
	\frac{\sqrt{3}\hat\omega_0}{\pi}\ln\frac{u}{1-u}$,
	$0<u<1$.
	
	At significance level $\alpha$, define the rejection region
	\[
	W=
	\left\{
	(z_{p+1},\ldots,z_m):
	\begin{array}{l}
		\text{there are more than }\alpha(m-p)\text{ indices }t,\ p+1\leq t\leq m,\\
		\text{such that }z_t<\Phi_0^{-1}\!\left(\frac{\alpha}{2}\right)
		\text{ or }z_t>\Phi_0^{-1}\!\left(1-\frac{\alpha}{2}\right)
	\end{array}
	\right\}.
	\]
	If $(\hat\eta_{0,p+1},\ldots,\hat\eta_{0,m})\in W$, then $H_0$ is rejected, indicating that the disturbance variance exhibits conditional heteroskedasticity. Otherwise, $H_0$ is not rejected, indicating that there is no evident conditional heteroskedasticity.
	
	After fitting the UGARCH$(q,s)$ model (\ref{eq:ugarch}), the residuals of the variance equation are
	\[
	\hat\eta_t
	=
	h_t-\hat\alpha_0
	-\sum_{i=1}^{q}\hat\alpha_i h_{t-i}
	-\sum_{j=1}^{s}\hat\beta_j\hat\sigma_{t-j}^{2},
	\qquad
	t=p+r+1,\ldots,m.
	\]
	Following the residual analysis for uncertain time-series models, the estimated mean and variance are
	\[
	\hat\theta
	=
	\frac{1}{m-p-r}\sum_{t=p+r+1}^{m}\hat\eta_t,
	\qquad
	\hat\omega^2
	=
	\frac{1}{m-p-r}\sum_{t=p+r+1}^{m}
	(\hat\eta_t-\hat\theta)^2.
	\]
	
	We assume that the estimated disturbance term $\hat{\eta}_t$
	follows a normal uncertainty distribution
	$N(\hat{\theta},\hat{\omega})$.
	To test this assumption, let $\theta_0=\hat{\theta}$
	and $\omega_0=\hat{\omega}$, and consider the hypotheses
	\[
	H_0:\theta=\theta_0\ \text{and}\ \omega=\omega_0
	\quad\text{versus}\quad
	H_1:\theta\neq\theta_0\ \text{or}\ \omega\neq\omega_0.
	\]
	The inverse uncertainty distribution of $N(\theta_0,\omega_0)$ is 
	$\Phi^{-1}(u)=\theta_0+\frac{\sqrt{3}\omega_0}{\pi}\ln\frac{u}{1-u}$, $0<u<1$.
	
	At significance level $\alpha$, define the rejection region
	\[
	W=
	\left\{
	(z_{p+r+1},\ldots,z_m):
	\begin{array}{l}
		\text{there are more than }\alpha(m-p-r)\text{ indices }t,\ p+r+1\leq t\leq m,\\
		\text{such that }z_t<\Phi^{-1}\!\left(\frac{\alpha}{2}\right)
		\text{ or }z_t>\Phi^{-1}\!\left(1-\frac{\alpha}{2}\right)
	\end{array}
	\right\}.
	\]
	If $(\hat\eta_{p+r+1},\ldots,\hat\eta_m)\in W$, then $H_0$ is rejected, and the fitted UGARCH$(q,s)$ model is not a good fit to the observed data. Otherwise, $H_0$ is not rejected, and the fitted UGARCH$(q,s)$ model is regarded as a good fit to the observed data.

	\subsection{Forecasting procedure}\label{sec:forecast}
	
	A key objective of time-series analysis is to characterize a series' dynamic behavior from observed data and forecast its future evolution. Because financial time-series volatility is generally time-varying, forecasting the next observation requires not only a forecast of the series itself but also consideration of dynamic changes in the disturbance variance. 	The UGARCH$(q,s)$ model  (\ref {eq:ugarch}) has been established above, this section develops a forecasting procedure for a time series with conditional heteroskedasticity.
	
	Let $\hat\alpha_0, \hat\alpha_1,\ldots, \hat\alpha_q, \hat\beta_1, \ldots, \hat\beta_s$ be the estimated UGARCH$(q,s)$ parameters. If the next-period uncertain disturbance follows $\eta_{m+1}\sim N(0,\hat\omega)$, the forecast uncertain variable for the next-period variance information is
	\[
	\hat h_{m+1}=\hat \alpha_0+\sum_{i=1}^{q} \hat\alpha_i h_{m+1-i}
	+\sum_{j=1}^{s}\hat \beta_j\hat\sigma_{m+1-j}^2+\eta_{m+1}.
	\]
	The forecast of the next-period disturbance variance is defined as the expected value of this uncertain variable:
	\[
	\hat\sigma_{m+1}^2=E[\hat h_{m+1}]
	=\hat\alpha_0+\sum_{i=1}^{q}\hat \alpha_i h_{m+1-i}
	+\sum_{j=1}^{s}\hat \beta_j\hat\sigma_{m+1-j}^2.
	\]
	Thus, the forecast of the next-period disturbance standard deviation is
	$
	\hat\sigma_{m+1}=\sqrt{\hat\sigma_{m+1}^2}.
	$
	
	The UAR model (\ref{UAR}) is $X_t=b_0+\sum_{j=1}^{p}b_j X_{t-j}+\epsilon_t$. For forecasting, we further assume that the next-period disturbance follows a normal uncertainty distribution with expected value $\hat e$ and conditional standard deviation $\hat\sigma_{m+1}$, i.e.,
	$
	\epsilon_{m+1}\sim N(\hat e,\hat\sigma_{m+1}),
	$
	where $\hat\sigma_{m+1}$ is obtained from the UGARCH variance recursion above. Therefore, the forecast uncertain variable for the next observation is
	\[
	\hat X_{m+1}=\hat b_0 +\sum_{j=1}^{p}\hat b_j X_{m+1-j}
	+N(\hat e,\hat\sigma_{m+1}).
	\]
	The forecast of the next observation is defined as the expected value of $\hat X_{m+1}$:
	$
	\hat\mu_{m+1}=\hat b_0+\sum_{j=1}^{p} \hat b_j X_{m+1-j}+\hat e.
	$
	Thus, $\hat X_{m+1}$ follows a normal uncertainty
	distribution $\mathcal{N}(\hat\mu_{m+1},\hat\sigma_{m+1})$.
	with uncertainty distribution function
	\[
	\hat\Phi_{m+1}(z)=\left[1+\exp\left(
	\frac{\pi(\hat\mu_{m+1}-z)}{\sqrt3\hat\sigma_{m+1}}\right)\right]^{-1}.
	\]
	For a confidence level $\alpha$, where $0<\alpha<1$, let
	$
	\hat c=\frac{\sqrt3\hat\sigma_{m+1}}{\pi}
	\ln\frac{1+\alpha}{1-\alpha}.
	$
	Then $\mathcal M\{\hat\mu_{m+1}-\hat c\leq\hat X_{m+1}\leq\hat\mu_{m+1}+\hat c\}\geq\alpha$. Therefore, a symmetric confidence interval for the next observation
	\( X_{m+1}\) at confidence level \(\alpha\) is
	\[
	\left[\hat\mu_{m+1}-\frac{\sqrt3\hat\sigma_{m+1}}{\pi}
	\ln\frac{1+\alpha}{1-\alpha},\quad
	\hat\mu_{m+1}+\frac{\sqrt3\hat\sigma_{m+1}}{\pi}
	\ln\frac{1+\alpha}{1-\alpha}\right].
	\]
	
	\section{Numerical Simulation}\label{sec4}
	
	The numerical study is organized into two parts. First, we examine the fitting performance of the proposed UGARCH model and its parameter estimation method. Second, we investigate the role of the lagged conditional variance term in describing variance persistence.

	\subsection{Fitting performance of the UGARCH model}
	
	To examine whether the proposed estimation method can recover the underlying UGARCH structure, we consider the UGARCH$(1,1)$ model
	\[
	h_t=\sigma_t^2+\eta_t,\qquad
	\sigma_t^2
	=
	\alpha_0+\alpha_1h_{t-1}
	+\beta_1\sigma_{t-1}^2,
	\]
	where $\eta_t$ is a centered uncertain disturbance satisfying $E[\eta_t]=0$ and $V[\eta_t]=\omega^2$. The true parameters are set to $\alpha_0=0.30$, $\alpha_1=0.30$, $\beta_1=0.40$, and $\omega=0.05$. Since $\alpha_1+\beta_1=0.70<1$, the stability condition is satisfied. The corresponding long-run conditional variance is $\alpha_0/(1-\alpha_1-\beta_1)=1$, and the initial conditional variance is therefore set to $\sigma_1^2=1$.
	
	A sample of $m=150$ observations is generated from the above UGARCH$(1,1)$ model under the framework of uncertainty theory. The disturbance sequence is constructed according to the normal uncertainty distribution $N(0,\omega)$, whose inverse uncertainty distribution is
	\[
	\Phi^{-1}(\alpha)
	=
	\frac{\sqrt{3}\omega}{\pi}
	\ln\frac{\alpha}{1-\alpha},
	\qquad 0<\alpha<1.
	\]
	The conditional variances and observations are then generated recursively from the specified model. The resulting disturbance sequence has a numerical mean approximately equal to zero and a numerical variance of $0.002449$, which is close to the theoretical value $\omega^2=0.0025$.
	
	The generated sample is fitted using the recursive constrained least-squares procedure in Section~3.2, with $\varepsilon=10^{-6}$ and convergence tolerance $10^{-8}$. The estimation procedure converges after 18 iterations. The parameter estimation results are reported in Table~\ref{tab:simfit}.
	
	\begin{table}[!htbp]
		\centering
		\caption{Simulation results for the UGARCH$(1,1)$ model.}
		\label{tab:simfit}
		\renewcommand{\arraystretch}{1.20}
		\begin{tabular*}{0.96\textwidth}{@{\extracolsep{\fill}}ccccc@{}}
			\toprule
			Quantity
			& True value
			& Estimated value
			& Absolute error
			& Relative error (\%) \\
			\midrule
			$\alpha_0$ & 0.30 & 0.3014 & 0.0014 & 0.48 \\
			$\alpha_1$ & 0.30 & 0.3265 & 0.0265 & 8.84 \\
			$\beta_1$ & 0.40 & 0.3723 & 0.0277 & 6.93 \\
			$\alpha_1+\beta_1$ & 0.70 & 0.6988 & 0.0012 & 0.17 \\
			Long-run variance & 1.00 & 1.0008 & 0.0008 & 0.08 \\
			$\omega$ & 0.05 & 0.0495 & 0.0005 & 1.06 \\
			\bottomrule
		\end{tabular*}
	\end{table}
	
	As shown in Table~\ref{tab:simfit}, the estimated values are generally close to their corresponding true values. In particular, the persistence coefficient is estimated as $0.6988$, compared with the true value $0.7000$, while the estimated long-run conditional variance is $1.0008$, compared with the true value $1.0000$. These results indicate that the recursive estimation procedure adequately recovers the main characteristics of the underlying UGARCH process.
	
	The fitting accuracy of the conditional variance is evaluated using the root mean squared error $\mathrm{RMSE}=\sqrt{\frac{1}{149}\sum_{t=2}^{150}(\hat{\sigma}_t^2-\sigma_t^2)^2}$ and the mean absolute error $\mathrm{MAE}=\frac{1}{149}\sum_{t=2}^{150}|\hat{\sigma}_t^2-\sigma_t^2|$. The resulting RMSE and MAE are $0.001961$ and $0.001278$, respectively, further confirming the accuracy of the fitted conditional variances under the present simulation setting.
	
	\FloatBarrier

	\subsection{Role of the lagged conditional variance term}
	
	We next examine the contribution of the lagged conditional variance term by comparing the UGARCH$(1,1)$ model with its UARCH$(1)$ counterpart. Since setting $\beta_1=0$ reduces the UGARCH$(1,1)$ model to the UARCH$(1)$ specification, the comparison provides a direct assessment of the contribution of $\beta_1\sigma_{t-1}^2$ to the variance dynamics.
	
	Both models are fitted to the same simulated sample. Their parameter estimates are reported in Table~\ref{tab:sim_uarch_ugarch_par}.
	
	\begin{table}[!htbp]
		\centering
		\caption{Parameter estimates of the UARCH$(1)$ and UGARCH$(1,1)$ models.}
		\label{tab:sim_uarch_ugarch_par}
		\renewcommand{\arraystretch}{1.25}
		\begin{tabular*}{0.90\textwidth}{@{\extracolsep{\fill}}lccc@{}}
			\toprule
			Model
			& $\hat{\alpha}_0$
			& $\hat{\alpha}_1$
			& $\hat{\beta}_1$ \\
			\midrule
			UARCH$(1)$
			& 0.6064
			& 0.3938
			& --- \\
			UGARCH$(1,1)$
			& 0.3014
			& 0.3265
			& 0.3723 \\
			\bottomrule
		\end{tabular*}
	\end{table}
	
\begin{figure}[!htbp]
		\centering
		\includegraphics[width=0.82\textwidth]{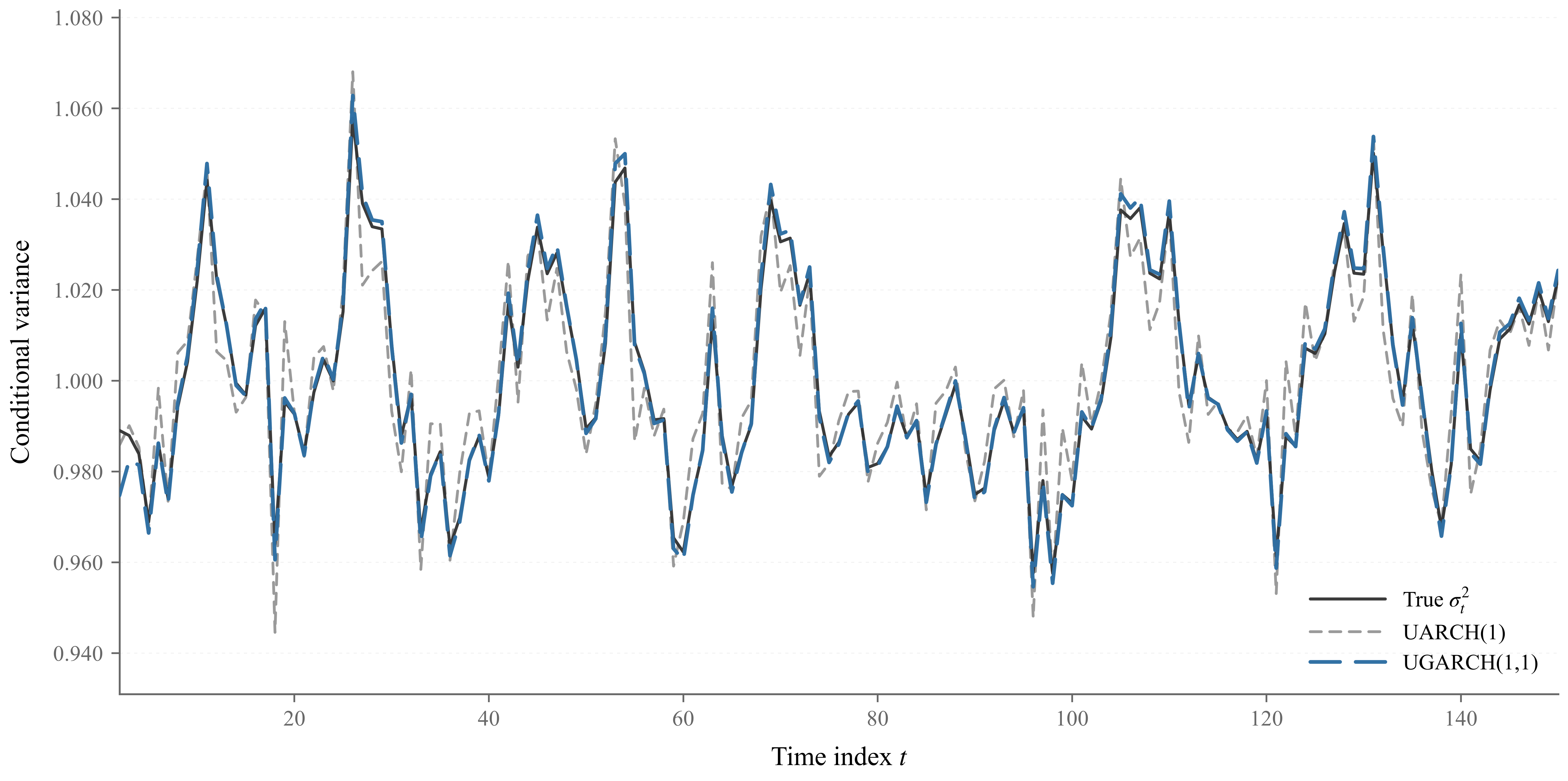}
		\caption{Comparison of the true and fitted conditional variances under the UARCH$(1)$ and UGARCH$(1,1)$ models.}
		\label{fig:sim_uarch_ugarch_fit}
	\end{figure}

	Figure~\ref{fig:sim_uarch_ugarch_fit} compares the fitted conditional variance paths from the two models with the true conditional variance. The UGARCH$(1,1)$ path follows the true conditional variance more closely throughout the sample, whereas the UARCH$(1)$ fit exhibits larger deviations, particularly during periods of relatively sharp fluctuations in the true conditional variance.
	
	For a quantitative comparison, we use the RMSE and MAE defined in the previous subsection. The results are reported in Table~\ref{tab:sim_uarch_ugarch_error}.
	
	\begin{table}[!htbp]
		\centering
		\caption{Conditional-variance fitting errors of the UARCH$(1)$ and UGARCH$(1,1)$ models.}
		\label{tab:sim_uarch_ugarch_error}
		\renewcommand{\arraystretch}{1.25}
		\begin{tabular*}{0.90\textwidth}{@{\extracolsep{\fill}}lccc@{}}
			\toprule
			Measure
			& UARCH$(1)$
			& UGARCH$(1,1)$
			& Reduction (\%) \\
			\midrule
			RMSE
			& 0.007864
			& 0.001961
			& 75.07 \\
			MAE
			& 0.006318
			& 0.001278
			& 79.77 \\
			\bottomrule
		\end{tabular*}
	\end{table}
	
	As shown in Table~\ref{tab:sim_uarch_ugarch_error}, introducing the lagged conditional variance term reduces the RMSE by approximately $75.07\%$ and the MAE by approximately $79.77\%$. Thus, under the present simulation setting, the term $\beta_1\sigma_{t-1}^2$ materially improves the representation of variance dynamics and persistence relative to the UARCH$(1)$ specification.
	
	\FloatBarrier

	\section{Empirical analysis of WTI crude oil future price}\label{sec5}
	The WTI crude oil futures closing-price data used in this study are
	obtained from Investing.com and are available at
	\url{https://cn.investing.com/commodities/crude-oil}. The prices are
	measured in US dollars per barrel. The 45 observations from May 21, 2025, to July 21, 2025, are used as the initial empirical sample.
	
	This period covers the geopolitical conflict in the Middle East in June
	2025, and the price series exhibits alternating episodes of pronounced and
	subdued volatility. It is therefore necessary to test further for
	conditional heteroskedasticity.
	
	\subsection{UAR model construction and estimation} 
	
\begin{figure}[htbp]
		\centering
		\includegraphics[width=0.82\textwidth]{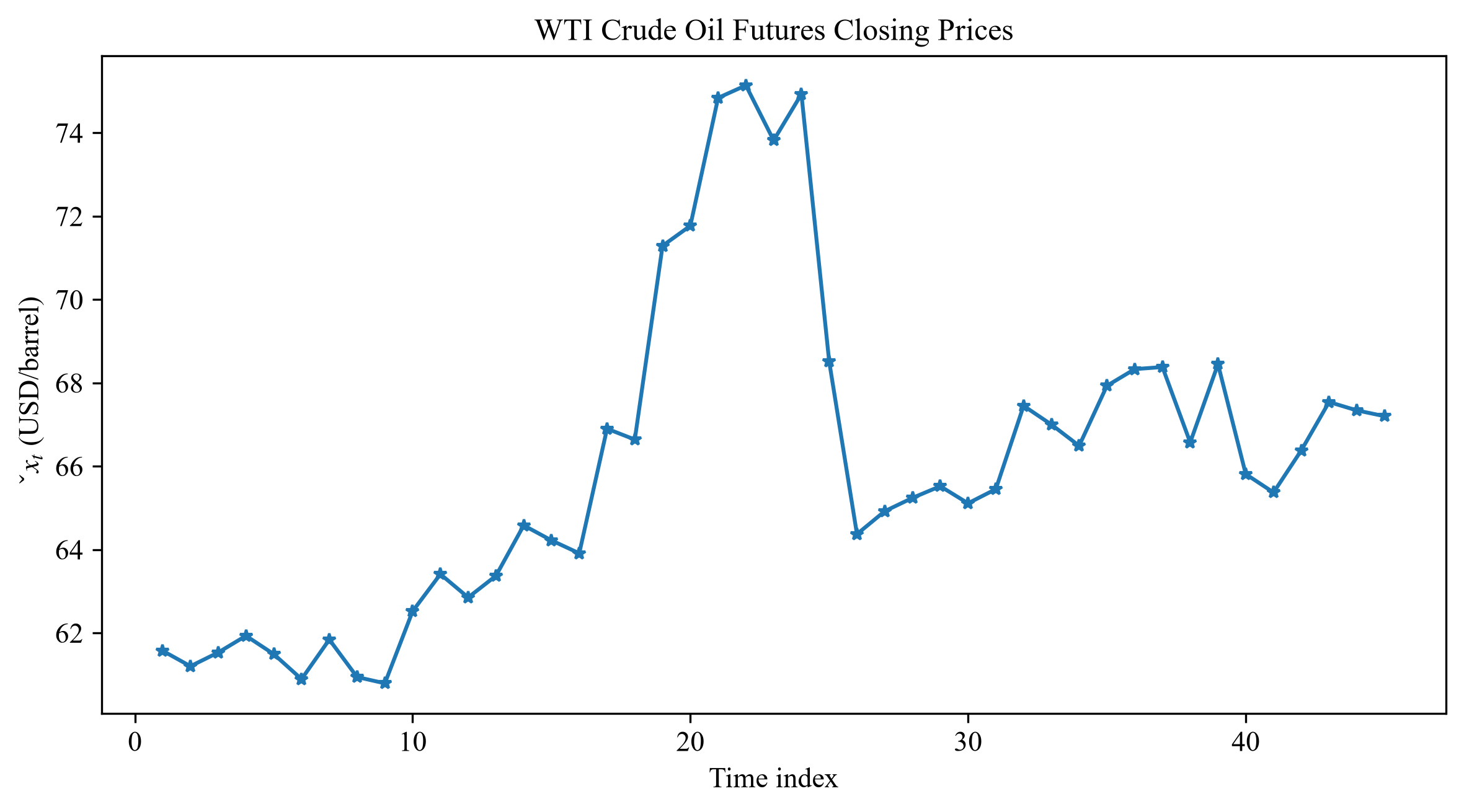}
		\caption{WTI crude oil futures closing-price series.}
		\label{fig:wti-price}
	\end{figure}

	Let \(x_t\) denote the closing price
	in period \(t\). The original price series is shown in
	Fig.~\ref{fig:wti-price}.
	Consider a UAR$(p)$ model (\ref{UAR}), formed by
	\[
	x_t
	=
	b_0+\sum_{j=1}^{p}b_jx_{t-j}+\varepsilon_t.
	\]
	Following the rolling-origin cross-validation procedure, set the initial
	training length to \(T=33\) and select the order from
	\(p=1,2,\ldots,5\). The corresponding average testing error is computed as
	\[
	\mathrm{ATE}_T(p)
	=
	\sum_{u=0}^{m-T-1}
	\frac{1}{m-T-u}
	\sum_{t=T+u+1}^{m}
	(x_t-\hat\mu_{t\mid T+u})^2.
	\]
	The average testing errors for the candidate orders are reported in
	Table~\ref{tab:uar-ate}.
	
	\begin{table}[htbp]
		\centering
		\caption{Average testing errors of the UAR$(p)$ models.}
		\label{tab:uar-ate}
		\renewcommand{\arraystretch}{1.25}
		\begin{tabular*}{\textwidth}
			{@{\extracolsep{\fill}}c c c c c c@{}}
			\toprule
			\(p\)&1&2&3&4&5\\
			\midrule
			\(\mathrm{ATE}_{33}(p)\)
			&\textbf{12.4301}
			&13.9344
			&14.7846
			&15.4883
			&16.0306\\
			\bottomrule
		\end{tabular*}
	\end{table}
	
	Table~\ref{tab:uar-ate} shows that
	\(\mathrm{ATE}_T(p)=12.4301\) at \(p=1\), which is the smallest value
	among the candidate orders. A UAR$(1)$ model is therefore selected.
	Least-squares estimation using the full sample gives
	$$
	x_t
	=
	7.9118+0.8821x_{t-1}+\varepsilon_t.
	$$

	The fitted residuals are computed period by period as
	$
	\hat\varepsilon_t
	=
	x_t-(7.9118+0.8821x_{t-1}).
	$
	The sample estimate of the disturbance expectation is
	$
	\hat e
	=
	\frac{1}{44}\sum_{t=2}^{45}\hat\varepsilon_t
	=
	-0.000000
	\approx0.
	$
	Hence, the squared centered residuals are
	$
	h_t
	=
	(\hat\varepsilon_t-\hat e)^2
	=
	\hat\varepsilon_t^{\,2}.
	$
	Selected results are presented in Table~\ref{tab:ht-result}.
	
	\begin{table}[htbp]
		\centering
		\caption{Selected squared centered residuals \(h_t\).}
		\label{tab:ht-result}
		\small
		\renewcommand{\arraystretch}{1.28}
		\begin{tabular*}{0.98\textwidth}
			{@{\extracolsep{\fill}}c c c c c c@{}}
			\toprule
			\(t\)&Date&\(x_t\)&Fitted value \(\hat\mu_t\)&
			Residual \(\hat\varepsilon_t\)&\(h_t\)\\
			\midrule
			2&2025-05-22&61.20&62.2217&\(-1.0217\)&1.0440\\
			3&2025-05-23&61.53&61.8954&\(-0.3654\)&0.1335\\
			4&2025-05-25&61.93&62.1865&\(-0.2565\)&0.0658\\
			5&2025-05-26&61.49&62.5393&\(-1.0493\)&1.1010\\
			17&2025-06-11&66.90&64.2858&2.6142&6.8339\\
			19&2025-06-13&71.29&66.6939&4.5961&21.1241\\
			21&2025-06-17&74.84&71.2190&3.6210&13.1116\\
			25&2025-06-23&68.51&74.0064&\(-5.4964\)&30.2102\\
			26&2025-06-24&64.37&68.3434&\(-3.9734\)&15.7880\\
			45&2025-07-21&67.20&67.3114&\(-0.1114\)&0.0124\\
			\bottomrule
		\end{tabular*}
	\end{table}
	
	\subsection{UGARCH-effect test and model fitting} 
	
	Under the constant-variance hypothesis \(H_0\), the conditional variance
	contains only a constant. The constant is first estimated from the 44
	squared centered residuals. Let the variance-equation disturbance under
	\(H_0\) be defined by
	$\hat\alpha_0=\frac{1}{44}\sum_{t=2}^{45} h_t=2.8319$ and 
	$\hat\eta_{0,t}=h_t-\hat\alpha_0.$
	The estimated scale parameter is
	$\hat\omega_0^2=\frac{1}{44}\sum_{t=2}^{45}\hat\eta_{0,t}^2
	=35.8226$ and 
	$\hat\omega_0=\sqrt{35.8226}=5.9852.$
	If the disturbance follows a normal uncertainty distribution with mean
	zero and scale parameter \(\hat\omega_0\), its inverse uncertainty
	distribution is
	\[
	\Phi_0^{-1}(u)
	=
	\frac{\sqrt3\hat\omega_0}{\pi}
	\ln\frac{u}{1-u},
	\qquad
	0<u<1.
	\]
	At significance level \(\alpha=0.05\), substituting
	\(u=0.025\) and \(u=0.975\) gives the 95\% confidence
	limits
	\[
	\Phi_0^{-1}(0.025)
	=
	-2.019827(5.9852)
	=
	-12.0891,
	\qquad
	\Phi_0^{-1}(0.975)
	=
	12.0891.
	\]
	According to the rejection region \(W\), the null hypothesis is
	rejected if more than \(\alpha(m-p)\) residuals fall outside these
	limits. Here, \(\alpha(m-p)=0.05\times44=2.2\).
	
	\begin{figure}[!htbp]
		\centering
		\includegraphics[width=0.82\textwidth]{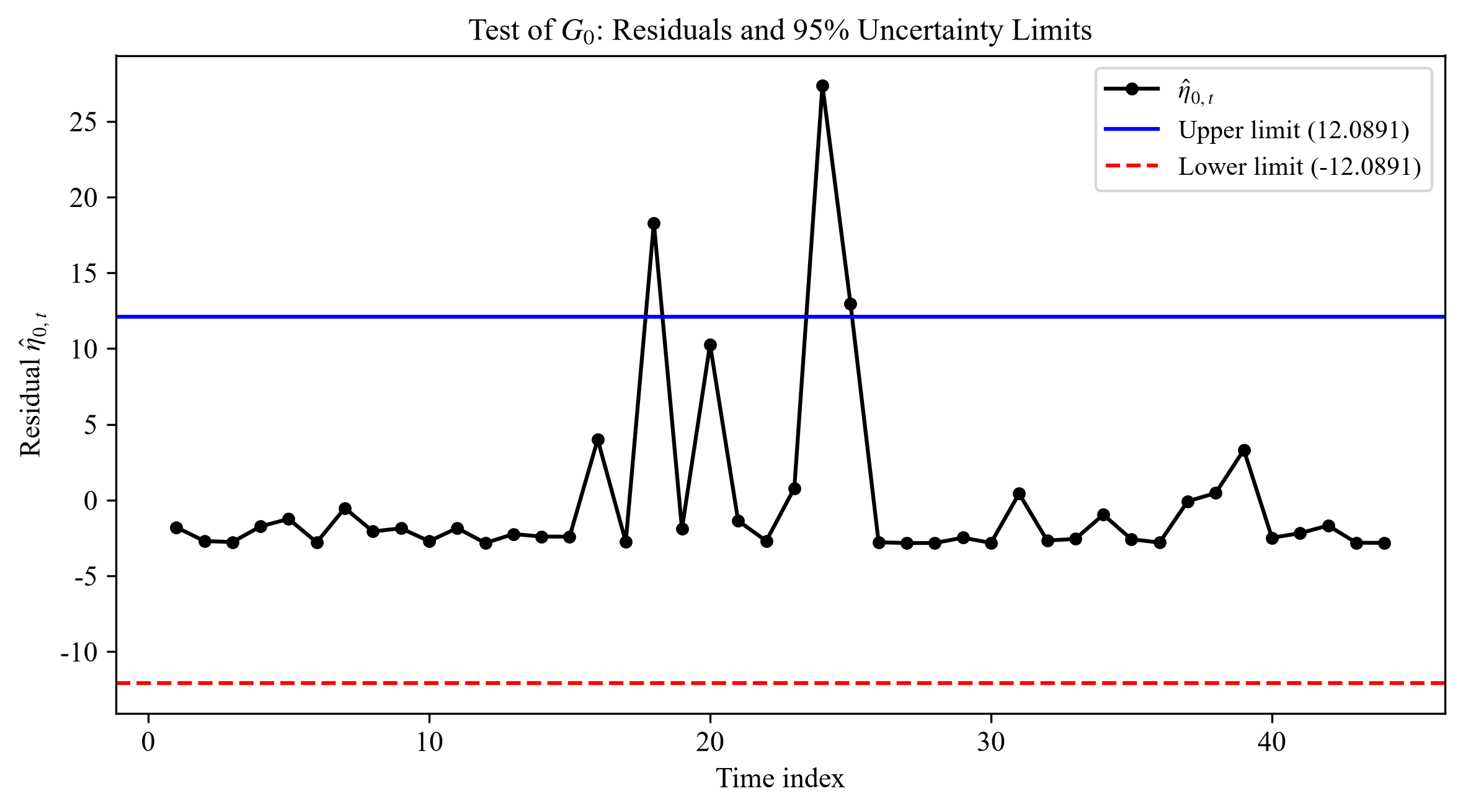}
		\caption{Test of \(H_0\): residuals and the 95\% uncertainty limits.}
		\label{fig:g0-test}
	\end{figure}
	
	As shown in Fig.~\ref{fig:g0-test}, three observations fall outside
	\([-12.0891,12.0891]\). Since \(3>2.2\), the residual vector belongs
	to \(W\). Thus, \(H_0\) is rejected at the 5\% significance level,
	providing evidence of a UGARCH effect.
	After rejecting \(H_0\), we fit the UGARCH$(q,s)$ model  (\ref {eq:ugarch}).
	Set the fixed training origin to \(T_h=30\) and perform cross-validation
	over the combinations \(q=1,2,3\) and \(s=0,1,2\).  
	
	\begin{table}[!htbp]
		\centering
		\caption{Order-selection results for the UGARCH$(q,s)$ models.}
		\label{tab:ugarch-order}
		\renewcommand{\arraystretch}{1.20}
		\small
		\begin{tabular*}{0.96\textwidth}
			{@{\extracolsep{\fill}}c c c c@{}}
			\toprule[1.1pt]
			$q$&\multicolumn{3}{c}{$\mathrm{ATE}_{T_h}(q,s)$}\\
			\cmidrule(lr){2-4}
			&$s=0$&$s=1$&$s=2$\\
			\midrule[0.6pt]
			1&5.988916&\textbf{5.778740}&6.204818\\
			2&6.107197&6.107197&6.107197\\
			3&6.574326&6.574326&6.574326\\
			\bottomrule[1.1pt]
		\end{tabular*}
	\end{table}
	
	Table~\ref{tab:ugarch-order} shows that the UGARCH$(1,1)$ model has an
	ATE of 5.7787, the smallest among all candidate models. The
	UGARCH$(1,1)$ model is therefore re-estimated using the full sample, giving
	\[
	\hat h_t
	=
	1.6346
	+0.1972 h_{t-1}
	+0.2345\hat\sigma_{t-1}^{\,2}
	+N(0,5.9083).
	\]
	Moreover,
	$\hat\alpha_1+\hat\beta_1
	=0.1972+0.2345=0.4317<1$,
	so the stationarity condition for the conditional variance process is
	satisfied. We then examine the adequacy of the estimated variance equation. Under
	the UGARCH$(1,1)$ model,
	\(h_t=\hat\sigma_t^2+\hat\eta_t\), so the fitted residual
	of the variance equation is
	\[
	\hat\eta_t
	=
	h_t-\hat\sigma_t^2
	=
	h_t-
	\left(
	1.6346+0.1972 h_{t-1}
	+0.2345\hat\sigma_{t-1}^2
	\right),
	\qquad
	t=3,\ldots,45.
	\]
	After excluding one recursive initial value, there are 43 effective
	residuals. The estimated scale parameter is
	$\hat\omega^2
	=\frac{1}{43}\sum_{t=3}^{45}\hat\eta_t^2
	=34.9081$ and
	$\hat\omega
	=\sqrt{34.9081}
	=5.9083.$
	The corresponding inverse uncertainty distribution is
	$\Phi^{-1}(u)
	=\frac{\sqrt3\hat\omega}{\pi}
	\ln\frac{u}{1-u},
	(0<u<1).$
	At \(\alpha=0.05\), the 95\% confidence limits are
	\[
	\Phi^{-1}(0.025)
	=
	-2.019827(5.9083)
	=
	-11.9338,
	\qquad
	\Phi^{-1}(0.975)
	=
	11.9338.
	\]
	According to the rejection region \(W\), the fitted model
	fails the residual goodness-of-fit test if more than
	\(\alpha(m-p-r)\) effective residuals fall outside these limits. Here,
	\(\alpha(m-p-r)=0.05\times43=2.15\).
	
	\begin{figure}[!htbp]
		\centering
		\includegraphics[width=0.82\textwidth]{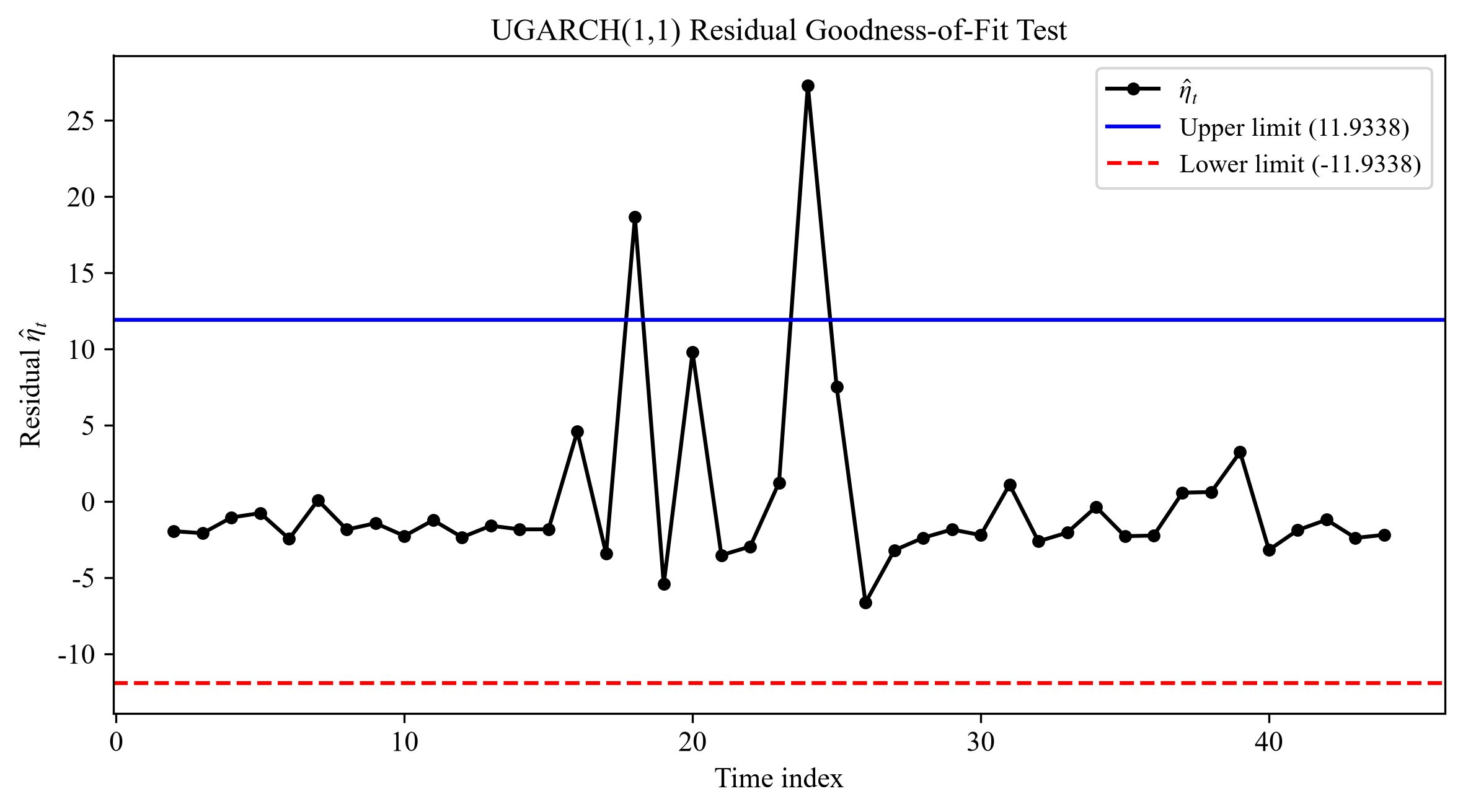}
		\caption{Fitted UGARCH residuals and the 95\% uncertainty limits.}
		\label{fig:g1-test}
	\end{figure}
	
	As shown in Fig.~\ref{fig:g1-test}, two observations fall outside
	\([-11.9338,11.9338]\). Since \(2<2.15\), the fitted residual vector
	does not belong to \(W\). Hence, the UGARCH$(1,1)$ model
	passes the residual goodness-of-fit test. Taken together, the rejection
	of \(H_0\) provides evidence of a UGARCH effect, while the residual
	goodness-of-fit test indicates that the fitted UGARCH$(1,1)$ model
	adequately describes the conditional variance dynamics of the sample.

	\FloatBarrier
	\subsection{One-step-ahead forecasting}
	
	Finally, the fitted UAR$(1)$--UGARCH$(1,1)$ model is used to forecast the
	WTI crude oil futures closing price on July 22, 2025.
	At the end of the sample, \(h_{45}=0.0124\) and
	\(\hat\sigma_{45}^2=2.2053\). The forecast uncertain variable for the
	next-period variance information is therefore
	\[
	\hat h_{46}
	=
	1.6346+0.1972(0.0124)+0.2345(2.2053)
	+N(0,5.9083)
	\approx
	2.1542+N(0,5.9083).
	\]
	The forecast of the next-period conditional variance is defined as the
	expected value of this uncertain variable. Because the disturbance has
	expected value zero,
	$
	\hat\sigma_{46}^2
	=
	E[\hat h_{46}]
	\approx
	2.1542.
	$
	Thus, the next-period conditional standard deviation is
	$
	\hat\sigma_{46}
	=
	\sqrt{\hat\sigma_{46}^2}
	\approx
	1.4677.
	$
	
	According to the estimated UAR$(1)$ mean equation, the uncertain variable
	for the next-period price is
	\[
	\hat X_{46}
	=
	7.9118+0.8821x_{45}+\varepsilon_{46},
	\qquad
	\varepsilon_{46}
	\sim
	N(0,1.4677).
	\]
	Substituting \(x_{45}=67.20\) and using the unrounded parameter
	estimates gives
	$
	\hat X_{46}
	\approx
	67.1879+N(0,1.4677).
	$
	Its expected value is the next-period point forecast:
	$
	\hat\mu_{46}
	=
	E[\hat X_{46}]
	\approx
	67.1879.
	$
	At confidence level \(\alpha\), since
	\(\hat X_{46}\sim
	N(\hat\mu_{46},\hat\sigma_{46})\), the radius of the
	symmetric prediction interval is
	$
	\hat c_\alpha
	=
	\frac{\sqrt3\hat\sigma_{46}}{\pi}
	\ln\frac{1+\alpha}{1-\alpha}.
	$
	For \(\alpha=0.90\) and \(\alpha=0.95\), respectively,
	$
	\hat c_{0.90}
	\approx
	2.3826,
	\hat c_{0.95}
	\approx
	2.9645.
	$
	Using the unrounded point and scale forecasts, the corresponding prediction
	intervals are
	$
	I_{0.90}
	=
	[64.8053,\ 69.5705]$,
	$
	I_{0.95}
	=
	[64.2233,\ 70.1524].
	$
	
	\begin{table}[htbp]
		\centering
		\caption{One-step-ahead forecasts of the WTI crude oil futures closing price.}
		\label{tab:forecast-result}
		\renewcommand{\arraystretch}{1.35}
		\setlength{\tabcolsep}{10pt}
		\begin{tabular*}
			{\textwidth}
			{@{\extracolsep{\fill}}l c c c@{}}\toprule
			Forecast quantity&Notation&Forecast\\
			\midrule
			Next-period conditional variance
			&\(\hat\sigma_{46}^2\)&\(2.1542\)\\
			Next-period conditional standard deviation
			&\(\hat\sigma_{46}\)&\(1.4677\)\\
			Next-period point forecast
			&\(\hat\mu_{46}\)&\(67.1879\)\\
			90\% prediction interval
			&\(I_{0.90}\)&\([64.8053,\ 69.5705]\)\\
			95\% prediction interval
			&\(I_{0.95}\)&\([64.2233,\ 70.1524]\)\\
			\bottomrule
		\end{tabular*}
	\end{table}
	
	The actual closing price on July 22, 2025, was 66.21 US dollars per barrel,
	which falls within both the 90\% and 95\% prediction intervals. This forecast
	is used as the first observation in the subsequent rolling out-of-sample
	evaluation.

	\subsection{Out-of-sample rolling forecast evaluation}
	
	Using the 45 observations from May 21 to July 21, 2025, as the initial
	training sample, 20 expanding-window one-step-ahead forecasts are generated
	for the period from July 22 to August 18, 2025. The model orders are fixed at
	UAR$(1)$--UGARCH$(1,1)$, and the parameters are re-estimated at each forecast
	origin. Selected forecasting results are presented in
	Table~\ref{tab:rolling-results}.
	
	\begin{table}[htbp]
		\centering
		\caption{Selected out-of-sample rolling forecasts.}
		\label{tab:rolling-results}
		\small
		\renewcommand{\arraystretch}{1.20}
		\begin{tabularx}{\textwidth}
			{c c c
				>{\centering\arraybackslash}X
				>{\centering\arraybackslash}X}
			\toprule
			Date & Actual price & Point forecast &
			90\% interval & 95\% interval
			\tabularnewline
			\midrule
			2025-07-22 & 66.2100 & 67.1879 &
			$[64.8053,\ 69.5705]$ & $[64.2233,\ 70.1524]$
			\tabularnewline
			2025-07-29 & 69.2100 & 66.7289 &
			$[64.2345,\ 69.2233]$ & $[63.6253,\ 69.8326]$
			\tabularnewline
			2025-07-31 & 69.2600 & 69.7150 &
			$[67.1624,\ 72.2675]$ & $[66.5390,\ 72.8909]$
			\tabularnewline
			2025-08-08 & 63.8800 & 64.2051 &
			$[61.8960,\ 66.5143]$ & $[61.3320,\ 67.0782]$
			\tabularnewline
			2025-08-14 & 63.9600 & 62.3830 &
			$[60.1070,\ 64.6589]$ & $[59.5511,\ 65.2148]$
			\tabularnewline
			2025-08-18 & 63.4200 & 63.1753 &
			$[60.7232,\ 65.6273]$ & $[60.1244,\ 66.2262]$
			\tabularnewline
			\bottomrule
		\end{tabularx}
	\end{table}
	
	For the 20 out-of-sample forecasts, the MAE is 1.0636 US dollars per barrel, and the MAPE is 1.6233\%. These results show that the point forecasts are
	generally close to the actual closing prices.
	
	All 20 actual prices fall within both the 90\% and 95\% prediction intervals.
	
	Overall, the rolling results show that the UGARCH$(1,1)$ model
	provides accurate short-term point forecasts and prediction intervals that
	cover all actual prices in the evaluation period.

\section{Conclusion}\label{sec6}

This paper mainly proposed a UGARCH model (\ref{eq:ugarch}) to describe
time-varying volatility and volatility persistence in uncertain time series,
an extension of the UARCH model (\ref{uarch}). By introducing lagged
conditional variances into the UARCH framework, the proposed model captures
the effects of both past volatility shocks and past conditional variances.
A sufficient condition for the mean stability of the conditional variance
process is established, and an iterative constrained least-squares procedure
is developed for parameter estimation. Cross-validation and uncertain
hypothesis tests are used for order selection and model assessment,
respectively. Numerical simulations demonstrate the parameter estimation
performance of the proposed method and show that UGARCH achieves smaller
conditional-variance fitting errors than UARCH under the considered settings.
The proposed model is applied to WTI crude oil futures closing prices, and
point forecasts and interval forecasts are conducted for the data.
The empirical results illustrate the applicability of the proposed model
in characterizing time-varying volatility and short-term forecasting.
Future research may consider more flexible extensions of the UGARCH model
(\ref{eq:ugarch}). For example, asymmetric UGARCH models may be developed
to capture different effects of positive and negative shocks on volatility.

\section*{Conflict of interest} 
The authors declare that they have no conflict of interest.

\section*{Data Availability }  
The data used in the motivating example are from https://cn.investing.com/commodities/crude-oil.

\section*{Funding}
The work was supported by the National Natural Science Foundation of China (12561047),  the Xinjiang Talent Development Fund (XJRC-2025-KJ-PY-KJLJ-108), and the 2025 Central Guidance for Local Science and Technology Development Fund (ZYYD2025ZY20).

\end{document}